\documentstyle[12pt,axodraw]{article}
\def\lesssim{\mathrel{\mathpalette\vereq<}}
\def\gtrsim{\mathrel{\mathpalette\vereq>}}

\def\lsim{\lesssim}
\def\gsim{\gtrsim}

\makeatletter
\def\vereq#1#2{\lower3pt\vbox{\baselineskip1.5pt \lineskip1.5pt
\ialign{$\m@th#1\hfill##\hfil$\crcr#2\crcr\sim\crcr}}}
\newcommand{\sla}[1]{{\raise.15ex\hbox{$/$}\kern-.57em #1}}
\makeatother

\begin{document}

\begin{titlepage}

\begin{flushright}
SNS-PH-01-12 \\
UCB-PTH-01/23 \\
LBNL-48257 \\
\end{flushright}

\vskip 1.5cm

\begin{center}
{\Large \bf  Models of Scherk-Schwarz Symmetry Breaking in 5D: \\ 
Classification and Calculability}

\vskip 1.0cm

{\bf
Riccardo Barbieri$^{a}$,
Lawrence J.~Hall$^{b,c}$,
Yasunori Nomura$^{b,c}$
}

\vskip 0.5cm

$^a$ {\it Scuola Normale Superiore and INFN, Piazza dei Cavalieri 7, 
                 I-56126 Pisa, Italy}\\
$^b$ {\it Department of Physics, University of California,
                 Berkeley, CA 94720, USA}\\
$^c$ {\it Theoretical Physics Group, Lawrence Berkeley National Laboratory,
                 Berkeley, CA 94720, USA}

\vskip 1.0cm

\abstract{The form of the most general orbifold breaking of gauge, 
global and supersymmetries with a single extra dimension is given. 
In certain theories the Higgs boson mass is ultraviolet finite 
due to an unbroken local supersymmetry, which is explicitly exhibited. 
We construct: a 1 parameter $SU(3) \times SU(2) \times U(1)$ theory 
with 1 bulk Higgs hypermultiplet, a 2 parameter $SU(3) \times SU(2) 
\times U(1)$ theory with 2 bulk Higgs hypermultiplets, and 
a 2 parameter $SU(5) \rightarrow SU(3) \times SU(2) \times U(1)$ 
theory with 2 bulk Higgs hypermultiplets, and demonstrate that 
these theories are unique.  We compute the Higgs mass and 
compactification scale in the $SU(3) \times SU(2) \times U(1)$ 
theory with 1 bulk Higgs hypermultiplet.}

\end{center}
\end{titlepage}

\section{Introduction} 
\label{sec:intro}

The origin of symmetry breaking, one of the key questions of particle
physics, is largely unknown. In four dimensions, symmetries can be
spontaneously broken by scalar fields, fundamental or composite. In
higher dimensional theories, a very different geometrical view is
possible: symmetries can be broken by the boundary conditions on a
compact space. While this idea has been known for many years 
\cite{Scherk:1979ta}, its early application was restricted to string
motivated theories with certain six dimensional compact spaces 
\cite{Candelas:1985en}.

It is remarkable that, until recently, there was no attempt to
discover the simplest extensions of the standard model, or the minimal
supersymmetric standard model (MSSM), in which symmetries, such as
supersymmetry, Peccei-Quinn symmetry and grand unified gauge symmetry,
were broken by this Scherk-Schwarz mechanism. While several such
theories now exist \cite{Antoniadis:1990ew, Antoniadis:1993fh, 
Pomarol:1998sd, Delgado:1999qr, Barbieri:2001vh, Arkani-Hamed:2001mi, 
Delgado:2001si, Kawamura:2000ev, Altarelli:2001qj, Hall:2001pg, 
Hebecker:2001wq, Barbieri:2001yz}, the question of 
their uniqueness is unknown, and is addressed in this paper.

We consider theories based on a single compact extra dimension. We
study the spacetime symmetries of this dimension in section 
\ref{subsec:spacetime}, and construct the most general form for the 
breaking of supersymmetry, global symmetry and gauge symmetry in 
section \ref{subsec:breaking}.

In section \ref{sec:symmetry} we exhibit the form of the local 5d gauge 
symmetry and supersymmetry which are unbroken by orbifolding. 
These unbroken symmetries depend on the location in the bulk -- 
for example they are different at the two fixed points of the orbifold. 
These unbroken symmetries are crucial since they dictate the form of 
both bulk and brane interactions. There has been considerable recent 
debate \cite{Ghilencea:2001ug} about whether the mass of the Higgs boson 
in certain theories of this type is finite. In section \ref{subsec:calc} 
we argue that these unbroken local 5d symmetries ensure that there are 
no quadratic ultraviolet divergences in the Higgs mass. 

In section \ref{sec:examples} we make a complete classification of 
theories with 5d local supersymmetry with one or two Higgs doublets 
in the bulk, with gauge group either $SU(3) \times SU(2) \times U(1)$ 
or $SU(5)$. There are very few such theories, and we briefly describe 
some possible locations for the matter fields.  Only a single 
$SU(3) \times SU(2) \times U(1)$ theory with a single bulk Higgs has 
been constructed in the literature \cite{Barbieri:2001vh}, and this 
theory is found to be an important special case of a 1 parameter family 
of such theories. We explore electroweak symmetry breaking in this 
family in section \ref{sec:deform}, with particular attention to the 
Higgs boson mass and the compactification scale.

The two Higgs theories, with gauge group $SU(3) \times SU(2) \times
U(1)$ and $SU(5)$, are shown in section \ref{sec:deform} to
each form unique two parameter families of models.
The form of $SU(5)$ breaking is unique, and the form for supersymmetry
breaking involves a single free parameter, and is therefore also 
highly constrained.  Conclusions are drawn in section \ref{sec:concl}.

\section{The Classification} 
\label{sec:class}

In this section we construct a classification of supersymmetric field
theories in 5 dimensions, where the physical space of the fifth
dimension is an orbifold of finite size. 

\subsection{5 dimensional spacetime}
\label{subsec:spacetime}

We begin by considering the fifth dimension to be the infinite line $R^1$.
What are the most general spacetime transformations acting on this
line, which can be used to compactify the spacetime by identifying points 
transforming into each other under these operations?  One of them is 
a translation ${\cal T} (2 \pi R)$ which induces $y \rightarrow y+2 \pi R$. 
When we identify the points connected by this transformation, that is 
$y + 2 \pi R$ with $y$, it compactifies $R^1$ to the circle $S^1 = R^1/T$. 
The other possibility is a parity ${\cal Z} (y_0)$ which reflects the 
line about $y=y_0$.  An identification using this operation, 
that identifies $-(y-y_0)$ with $y-y_0$, produces an orbifold 
which is the half line, $R^1/Z_2$. This identification involves 
the choice of a special point, $y_0$, which is a fixed point under 
the transformation. This parity alone does not compactify
the space. No further independent spacetime identifications can be
made on the line (if there are several commensurate translations, we
take ${\cal T}$ to be the one of lowest $R$).  In this paper, we are 
interested in the case that both translation and parity identifications 
are made. In this case the physical space can be taken to be 
$0 \leq (y-y_0) \leq \pi R$, corresponding to the orbifold $S^1/Z_2$. 

Let $\varphi$ be a column vector representing all fields of the theory. 
The action of these transformations on the fields can be written as
\begin{eqnarray}
  {\cal T}(2\pi R) [\varphi(y)] &=& T^{-1} \varphi(y+2 \pi R), 
\label{eq:calT} \\
  {\cal Z}(0) [\varphi(y)] &=& Z \varphi(-y),
\label{eq:calZ}
\end{eqnarray}
where we have chosen $y_0 = 0$. Acting with ${\cal Z}$ twice produces 
the identity, so that this is a $Z_2$ transformation, with $Z^2 = 1$. 
An identification under these operations are made by imposing the 
conditions ${\cal T}(2 \pi R) [\varphi(y)] = \varphi(y)$ 
and ${\cal Z}(0) [\varphi(y)] = \varphi(y)$, that is
\begin{eqnarray}
  \varphi(y+2 \pi R) &=& T \varphi(y),
\label{eq:Taction} \\
  \varphi(-y) &=& Z \varphi(y).
\label{eq:Zaction}
\end{eqnarray}
This identification makes sense only when the bulk action is invariant 
under the operations $\varphi(y) \rightarrow {\cal T}(2 \pi R) 
[\varphi(y)]$ and $\varphi(y) \rightarrow {\cal Z}(0) [\varphi(y)]$, 
since otherwise physics is not the same on all equivalent pieces 
of the line of length $\pi R$.

The simultaneous imposition of both ${\cal T}(2\pi R)$ and ${\cal Z}(0)$ 
is not automatically consistent, because the spacetime motion induced 
by ${\cal T}(2\pi R) {\cal Z}(0)$ is identical to that induced by
${\cal Z}(0) {\cal T}^{-1}(2\pi R)$. Consistency therefore requires that 
the field transformation is the same nomatter which choice is made; thus
we require $TZ = Z T^{-1}$, or
\begin{equation}
  ZTZ = T^{-1}.
\label{eq:TZconsistency}
\end{equation}
Thus the most general spacetime symmetries can be taken to be a reflection
$y \rightarrow -y$, under which the fields transform as a $Z_2$, and a
translation, under which the fields transform as Eq.~(\ref{eq:calT}) with 
any symmetry $T$ of the action, as long as Eq.~(\ref{eq:TZconsistency}) 
is satisfied.

\begin{figure}
\begin{center}
\begin{picture}(160,180)(-80,-90)
  \CArc(0,0)(60,0,360)
  \DashLine(-80,0)(80,0){2} \Text(-90,0)[r]{$P$}
  \DashLine(0,-80)(0,80){2} \Text(0,90)[b]{$P'$}
  \CArc(0,0)(61,180,205) \CArc(0,0)(59,180,205)
  \Line(-54,-26)(-52,-20) \Line(-54,-26)(-60,-23) \Text(-48,-22)[l]{$y$}
  \CArc(0,0)(61,270,295) \CArc(0,0)(59,270,295)
  \Line(26,-54)(20,-52) \Line(26,-54)(23,-60) \Text(25,-46)[b]{$y'$}
  \Vertex(-60,0){3} \Text(-65,10)[br]{$O$}
  \Vertex(0,-60){3} \Text(-5,-70)[ur]{$O'$}
  \Vertex(60,0){3}  \Text(65,-10)[ul]{$O$}
  \Vertex(0,60){3}  \Text(5,70)[bl]{$O'$}
\end{picture}
\caption{A diagrammatic representation of  ${\cal Z}(0)$ and
${\cal Z}(\pi R)$ as reflections about $y=0$ and $y'=0$, with $y'=y-\pi R$.}
\label{Fig_space}
\end{center}
\end{figure}
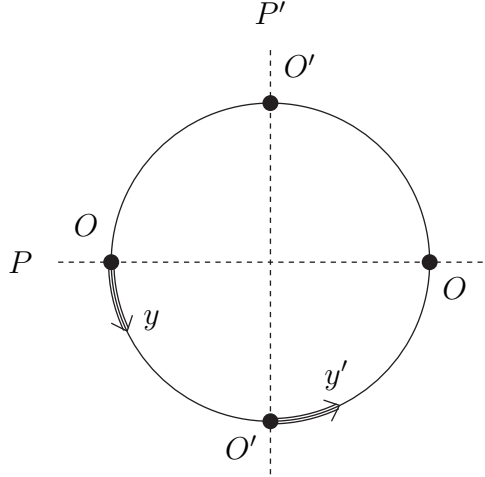

The compound transformation ${\cal T}(2 \pi R) {\cal Z}(0)$,
induces the spacetime motion $y - \pi R \rightarrow -(y - \pi R)$, which
is a reflection about the point $y=\pi R$. Its action on the
fields is $Z' = TZ$, and from Eq.~(\ref{eq:TZconsistency}) we discover
that $Z'^2=1$, so that ${\cal T}(2 \pi R) {\cal Z}(0) = {\cal Z}(\pi
R)$ is a reflection about $\pi R$ which also induces a $Z_2$ transformation
on the fields. One can choose to describe the compactification in
terms of the identifications  ${\cal Z}(0)$ and ${\cal T}(2\pi R)$ or
equivalently by  ${\cal Z}(0)$ and ${\cal Z}(\pi R)$; the orbifolds 
$S^1/Z_2$ and $R^1/(Z_2, Z_2')$ are equivalent \cite{Barbieri:2001vh}.
While the physical space is the line segment $0 < y < \pi R$, we have
found it convenient to assemble four such equivalent neighboring
segments into a circle of circumference $4 \pi R$, as shown in Figure 
\ref{Fig_space}. The utility of this construction is to provide a 
diagrammatic view of ${\cal Z}(0)$ and ${\cal Z}(\pi R)$ as reflections 
about orthogonal axes with fixed points $O$ and $O'$.

In general $Z$ and $Z'$ do not commute. In the special case that they
do, $T^2 = 1$, so that $T$ is also a $Z_2$ transformation. Acting
twice with ${\cal T}(2\pi R)$ induces a complete revolution of the circle
of Figure \ref{Fig_space}, so that, in this commuting case, the 
eigenfunctions of Eqs.~(\ref{eq:Taction}, \ref{eq:Zaction}) are single 
valued on this circle:
\begin{eqnarray}
T &=& +1 \;\; \left\{\begin{array}{ll}
		(+,+): \;\; & \cos \left[ ny / R \right] \\
		(-,-): \;\; & \sin \left[ ny / R \right] \nonumber \\
		\end{array} \right. \\
T &=& -1 \;\; \left\{\begin{array}{ll}
		(+,-): \;\; & \cos \left[ (n+1/2)y / R \right] \\
		(-,+): \;\; & \sin \left[ (n+1/2)y / R \right] \\
		\end{array} \right.
\label{eq:commuting.eigenfunctions}
\end{eqnarray}
where $(\pm, \pm)$ refer to the $(Z,Z')$ parities.

Given a specific field content of a theory, a complete list of the
possible forms for $Z$ and $Z'$ can be obtained. As an illustrative
example, consider a theory with $N$ complex scalars, assembled into a
vector $\phi$. A basis can be chosen such that $Z=P$ is a diagonal
matrix. However, in this basis $Z'$ is in general non-diagonal:
\begin{equation}
  Z'(\alpha_i) = U P' U^\dagger,
\label{eq:Z'1}
\end{equation}
where $P'$ is diagonal. The $N \times N$ unitary matrix $U(\alpha_i)$
describes the relative orientation of the field bases which
diagonalize $Z$ and $Z'$, 
and depends on a set of continuous parameters $\alpha_i$.

The number of physical parameters $\alpha_i$ is less than $N^2$, and
depends on the numbers of positive and negative eigenvalues in $P$ and
$P'$. Two cases will be of particular importance to us. 
If either $P$ or $P'$ is proportional to the identity, then $U$
can be rotated away, and there is no need to introduce any $\alpha_i$ 
parameters. Next consider the case of $N=2$. The only non-trivial case 
is when neither $P$ or $P'$ is proportional to the unit matrix: 
$P=P'=\sigma_3$. A general $U$ matrix would have the form
$U=\exp(i \sum_{i=0}^{3} \alpha_i \sigma_i)$, where $\sigma_0$ is the 
unit matrix and $\sigma_{1,2,3}$ are the Pauli spin matrices. 
However, the parameters $\alpha_{0,3}$ drop out of $Z'$, while a basis 
rotation which preserves $Z=\sigma_3$ allows $\alpha_1$ to be 
rotated away. Hence the only non-trivial $2 \times 2$ case is 
described by a single parameter $\alpha$:
\begin{equation}
  Z=\sigma_3, \qquad\qquad Z'(\alpha) = e^{\pi i \alpha \sigma_2}
  \sigma_3 e^{-\pi i \alpha \sigma_2}. 
\label{eq:ZZ'}
\end{equation}
The description in terms of $(Z,T)$ is somewhat simpler, since $T=
e^{2 \pi i \alpha \sigma_2} = R(2\pi\alpha)$, the $2 \times 2$ rotation 
matrix for angle $2\pi\alpha$. In this case the field $\phi$ can be 
expanded in a set of Kaluza-Klein (KK) eigenfunctions of 
Eqs.~(\ref{eq:Taction}, \ref{eq:Zaction}):
\begin{equation}
  \phi(x,y) = R \left( \alpha {y \over R} \right)
    \sum_{n=0}^\infty \left( \begin{array}{l}
	\cos \left[ ny / R \right] \phi_{+n} (x) \\
	\sin \left[ ny / R \right] \phi_{-n} (x)
    \end{array} \right)
  = \sum_{n=-\infty}^\infty \left( \begin{array}{ll}
        \cos \left[ (n + \alpha)y / R \right]  \\
        \sin \left[ (n + \alpha)y / R \right] 
    \end{array} \right)  \phi_n (x),
\label{eq:N=2eigenfunctions}
\end{equation}
where $\phi_n$ is given by 
\begin{equation}
  \phi_n(x) = 
    \left\{\begin{array}{ll}
      \frac{1}{2} (\phi_{+n}(x) + \phi_{-n}(x)) \quad & {\rm for}\;\; n>0 \\
      \phi_{+0}(x)                              \quad & {\rm for}\;\; n=0 \\
      \frac{1}{2} (\phi_{+n}(x) - \phi_{-n}(x)) \quad & {\rm for}\;\; n<0.
    \end{array} \right.
\end{equation}
The special cases $\alpha = 0$ ($1/2$) give $T=\sigma_0$ ($-\sigma_0$), 
so that $T^2 = 1$. In these cases $Z$ and $Z'$ commute, so that the above 
eigenfunctions Eq.~(\ref{eq:N=2eigenfunctions}) reduce to 
Eq.~(\ref{eq:commuting.eigenfunctions}).

\subsection{General form for orbifold symmetries} 
\label{subsec:breaking}

We consider $N=1$ supersymmetric gauge theories in 5d with gauge 
group $G$. The vector multiplet contains components ${\cal V} = 
(A^M,\lambda,\lambda',\sigma)$ and the theory contains a set of
hypermultiplets with components ${\cal H} = (\phi, \phi^{c \dagger}, \psi,
\psi^{c \dagger})$. There may be multiple copies of a hypermultiplet of 
given gauge charge, and therefore, since supersymmetry allows only kinetic
terms in the bulk, the bulk Lagrangian can possess some flavor symmetry
$H$. From the 4d viewpoint the theory possesses two supersymmetries,
with transformation parameters $\Xi = (\xi_1(y), \xi_2(y))$, which 
we take to be local transformations.
The bulk Lagrangian possesses a global $SU(2)_R$ symmetry under which 
$\Xi = (\xi_1, \xi_2)$, $\Lambda = (\lambda, \lambda')$ and $\Phi =
(\phi, \phi^{c \dagger})$ form doublets.

The symmetry ${\cal Z}(0)$ induces $y \rightarrow -y$ and the
supersymmetric kinetic terms then force relative signs for the
parities $P$ of the components inside ${\cal V}$ or
${\cal H}$. In particular one discovers that, from the 4d viewpoint,
${\cal Z}$ necessarily breaks $N=2$ supersymmetry to $N=1$ supersymmetry. 
The 5d supersymmetric multiplets are then conveniently assembled into 
4d supersymmetric multiplets with the $P$ charges: 
${\cal V} = (V(+),\Sigma(-))$ and ${\cal H} = (H(+), H^{c \dagger}(-))$, 
where $V(A^\mu, \lambda)$ is a 4d vector multiplet,
whereas $\Sigma(\sigma + i A^5, \lambda')$, $H(\phi, \psi)$ and 
$H^c(\phi^c, \psi^c)$ are chiral multiplets. This action of ${\cal Z}$ 
within an $N=2$ multiplet we define as the set of charges
$\Sigma_3$. However, this does not give the complete action of
${\cal Z}$. There may be different overall phase rotations for 
different hypermultiplets, $P_H$. Finally, even within a hypermultiplet,
${\cal Z}$ can act differently on different components of an
irreducible gauge multiplet. We label this by $P_G$, which we take to be 
an element of the gauge group.  If $P_G$ is the unit matrix there is 
no gauge symmetry breaking, otherwise there is. Hence we write
\begin{equation}
  Z = \Sigma_3 \otimes P_H \otimes P_G.
\label{eq:Z}
\end{equation}

The most general possibilities for $P_H$ and $P_G$ are given by 
$P_H^2 = P_G^2 = \pm 1$ so that $Z^2 = (P_H \otimes P_G)^2 = 1$ 
(not just $P_H^2 = P_G^2 = 1$).  An example of the case with 
$P_H^2 = P_G^2 = -1$ is provided by $G = SU(2)$ with two iso-doublet 
hypermultiplets ${\cal H}_\pm = (H_\pm, H_\pm^{c\dagger})$.
In this case, to have a non-trivial boundary condition in the gauge 
space ($P_G \neq 1$), we have to take $P_G = i \sigma_3$ in the 
space of fundamental representation. ($P_G = \sigma_3$ is not an 
element of $SU(2)$.)  Therefore, $P_G^2 = -1$.  Then, to have 
$Z^2 = 1$, we also have to assign $P_H^2 = -1$ for ${\cal H}_\pm$, 
for instance, as $H_\pm \rightarrow \mp i H_\pm$ and 
$H_\pm^c \rightarrow \pm i H_\pm^c$.  The combined transformation, 
$P_H \otimes P_G$, is written as $H_\pm \rightarrow \pm \sigma_3 H_\pm$ 
and $H_\pm^c \rightarrow \mp \sigma_3 H_\pm^c$, which cannot be 
reproduced in terms of the parity matrices $P_H$ and $P_G$ satisfying 
$P_H^2 = P_G^2 = 1$, since with $P_H^2 = P_G^2 = 1$ the induced 
transformation is always the same for $H_\pm$ and $H_\pm^c$. Indeed, 
in this $SU(2)$ case, the transformations for $H_\pm$ and $H_\pm^c$ 
must be opposite under $P_H \otimes P_G$, in spite of the fact that 
$H_\pm$ and $H_\pm^c$ belong to the same representation, ${\bf 2}$, 
under the $SU(2)$. Similar situations also occur, for instance, 
in the case of $G = SO(10)$ with vector representations.
However, all explicit models we discuss in this paper are described by 
$P_H^2 = P_G^2 = 1$, because the gauge breaking considered in these 
models are only $SU(5) \rightarrow SU(3) \times SU(2) \times U(1)$. 
Therefore, we call $P_H$ and $P_G$ as parity matrices, but it should 
be understood that the eigenvalues for them can be $\pm i$ in general.

The argument for ${\cal Z}(0)$ applies identically to the symmetry
${\cal Z}(\pi R)$, but the basis which diagonalizes $Z'$ is in
general different from that which diagonalizes $Z$. Choosing $Z$
diagonal, we immediately find that $Z'$ must take the form
\begin{equation}
  Z' = U \Sigma_3 U^\dagger \otimes V P'_H V^\dagger \otimes W P'_G W^\dagger,
\label{eq:Z'}
\end{equation}
where $P'_{H,G}$ are diagonal matrices with 
$P_{H}^{\prime 2} = P_{G}^{\prime 2} = \pm 1$, 
$U,V$ are unitary matrices, and $W$ is an appropriate matrix making 
the action invariant under the operation ${\cal Z}(\pi R)$ (for instance, 
$W$ is a unitary (orthogonal) matrix for $G = SU(n)$ ($SO(n)$)). 
Note that $U$ is an element of
$SU(2)_R$. Thus the action of $\Sigma_3$ and $ U \Sigma_3 U^\dagger$
is the same on the $SU(2)_R$ singlets: 
$(\psi, \psi^c) \rightarrow (\psi, -\psi^c)$ and 
$(A^\mu, \sigma + i A^5) \rightarrow (A^\mu, -(\sigma + i A^5))$, but 
differs on the $SU(2)_R$ doublets $\Lambda$ and $\Phi$. 
From the discussion preceding Eq.~(\ref{eq:ZZ'}), there is no loss of 
generality in taking $\Sigma_3 = \sigma_3$ and 
$U= e^{\pi i \alpha \sigma_2}$ acting on the $SU(2)_R$ doublet space. 
On the other hand, the forms for $V$ and $W$ are highly model dependent. 
For example, if the flavor group $H=U(1)^N$, then $V=1$.
If $H$ contains $SU(2)$ factors, then in the corresponding $2 \times 2$ 
blocks, $V$ is either $\sigma_0$ or $e^{\pi i \beta \sigma_2}$, 
depending on whether the two hypermultiplets of the same gauge charge 
have equal or opposite $P$ parities.

Throughout this paper, we consider theories with local supersymmetry in 
the bulk. However the above classification is unchanged in the 
non-supersymmetric case, as long as the action of $\Sigma_3$ is reinterpreted. 
It acts as $(A^\mu, A^5) \rightarrow (A^\mu, -A^5)$ and 
$(\psi, \eta) \rightarrow (\psi, -\eta)$, where $\psi$ 
and $\eta$ are the components of any 5d Dirac fermion.

\section{Symmetries and Symmetry Breaking}
\label{sec:symmetry}

Every non-trivial entry in Eqs.~(\ref{eq:Z}, \ref{eq:Z'}) causes
symmetry breaking.  One of the two supersymmetries is broken by
$\Sigma_3$, and the other is broken by a non-trivial $U$. The flavor
symmetry $H$ is broken by $P_H, P_H', V$ and the gauge symmetry $G$
by $P_G,P_G',W$. 

\subsection{Supersymmetry breaking}
\label{subsec:susy}

If the extra dimension is not compactified, the theory is
invariant under the local supersymmetry transformations $\delta
\psi(x,y) =  \Xi^T (x,y)\, \sla{\partial} \Phi, \cdots$ with $\Xi(x,y) 
= (\xi_1(x,y), \xi_2(x,y) )$ an arbitrary function of
spacetime. Compactification with the orbifold boundary conditions of
Eqs.~(\ref{eq:Z}, \ref{eq:Z'}) reduces the set of local supersymmetry
transformations. The action of $Z$ and $Z'$ in $SU(2)_R$ space is
given by Eq.~(\ref{eq:ZZ'}) so that the theory is invariant under
supersymmetry transformations with the form of
Eq.~(\ref{eq:N=2eigenfunctions}) 
\begin{equation}
  \Xi(x,y) = \left( \begin{array}{l}
		\xi_1(x,y) \\
		\xi_2(x,y)
	     \end{array} \right)
  = R \left( \alpha {y \over R} \right)
    \sum_{n=0}^\infty \left( \begin{array}{l}
	\cos [ny / R]\, \xi_{+n} (x) \\
	\sin [ny / R]\, \xi_{-n} (x)
    \end{array} \right).
\label{eq:localsusy}
\end{equation}
Although this is a significant restriction, the theory still possesses
local 5d supersymmetry. From the low energy viewpoint, the number of
4d supersymmetries is the number of independent modes of
Eq.~(\ref{eq:localsusy}) having $\Xi = \Xi(x)$ independent of $y$. There
is at most a single zero-mode, since the action of both $Z$ and $Z'$
necessarily involves $\sigma_3$. In fact, an unbroken 4d supersymmetry
only results if $\alpha = 0$, in which case it is the mode:
\begin{equation}
  \Xi(x) =  \left( \begin{array}{c}
		\xi_{+0}(x) \\
		0
	    \end{array} \right).
\label{eq:susyzeromode}
\end{equation}
For $\alpha \neq 0$, Eq.~(\ref{eq:localsusy}) has no zero-mode and hence
no 4d supersymmetry survives into the infrared. One supersymmetry is
broken by $1/R$, the other by $\alpha /R$. For any $\alpha \neq 0$, 
the $n$th mode is proportional to
$(\cos [(n+\alpha)y/R], \sin [(n + \alpha)y/R]) \xi_n(x)$, and has an 
axis which rotates with $y$. For $\alpha = 1/2$,
the 4d supersymmetries on the branes at $y=0$ and $y = \pi R$ 
are orthogonal. 

\subsection{Global symmetry breaking}
\label{subsec:global}

The global symmetry $H$ arises from a repetition of hypermultiplets
with the same gauge quantum numbers. If $H$ is generated by
$T^\alpha$, then the identification by ${\cal Z}(0)$ breaks those 
generators for which $[P_H, T^\alpha] \neq 0$. Similarly an 
identification by ${\cal Z}(\pi R)$ breaks those generators having 
$[V P_H' V^{\dagger}, T^\alpha] \neq 0$. 
The unbroken global group $H'$ is generated by the set of 
generators which commutes with both $P_H$ and $V P_H' V^{\dagger}$.

\subsection{Gauge symmetry breaking}
\label{subsec:gauge}

With a non-compact fifth dimension, the theory is invariant under
gauge transformations of $G$: $\delta A^{aM} = \partial^M
\epsilon^a(y) + \cdots$, with arbitrary gauge transformation 
parameters $\epsilon^a(y)$. Compactification with non-trivial
$P_G$ implies that the gauge fields split up into two sets $A^a
= (A^{a_+}, A^{a_-})$, which are $(+,-)$ under $y \rightarrow -y$.
The $+$ modes have generators which commute with $P_G$:  
$[P_G, T^{a_+}] = 0$. 
Hence the compactified theory possesses only a restricted gauge
symmetry with gauge parameters constrained to satisfy \cite{Hall:2001pg}
\begin{equation}
  \epsilon^{a_\pm}(-y) = \pm \epsilon^{a_\pm}(y).
\label{eq:epsilon}
\end{equation}
On making a KK mode expansion, $A^{a_+}$ have zero-modes while 
$A^{a_-}$ do not, so that the low energy 4d gauge group is $G_+$, 
generated by $T^{a_+}$. We frequently say that compactification 
using the parities $P_G$ has induced the gauge symmetry breaking 
$G \rightarrow G_+$. An alternative viewpoint is that the 
theory on the compact space possesses a restricted
set of gauge transformations, Eq.~(\ref{eq:epsilon}), which are not
broken.  They do not include zero-mode transformations of $G/G_+$. 

A precisely analogous argument applies for the gauge symmetry breaking 
induced by $P_G'$: $G \rightarrow G'_+$.  If $W = 1$, so that 
$P_G$ and $P_G'$ are simultaneously diagonalizable, then the zero-mode 
gauge bosons correspond to the generators which commute with both 
$P_G$ and $P_G'$, and are therefore $(+,+)$ modes. 
The lightest gauge boson mode for other generators have masses of
order $1/R$. These modes are either $(+,-), (-,+)$ or $(-,-)$. For
$W=1$, the mode eigenfunctions are given by
Eq.~(\ref{eq:commuting.eigenfunctions}). 

In the case that $W(\gamma)$ has a non-zero Euler 
angle, $\gamma$, further gauge symmetry generators are broken 
$[W P_G' W^{\dagger}, T^a] \neq 0$, with some 
previously massless gauge bosons acquiring mass $\gamma /R$. 
In this case the KK modes of the local gauge transformations have
forms which depend on the continuous parameter $\gamma$.
Thus in general the total structure of gauge symmetry breaking, 
$G \rightarrow G'$, is very rich.

\subsection{The brane action}
\label{subsec:brane}

The action has both bulk and brane contributions
\begin{equation}
  S = \int d^4x \; dy \; 
	\left[ {\cal L}_5 + \delta(y) {\cal L}_4
	  + \delta(y- \pi R) {\cal L}_4'  \right].
\label{eq:action}
\end{equation}
The form of the bulk action is very tightly constrained by the
unbroken local  5d gauge and supersymmetry transformations discussed
above. What interactions are allowed on the branes at $y= 0, \pi R$?

The constraints imposed by the local symmetries are found in the 
following way: the brane actions ${\cal L}_4, {\cal L}_4'$ are the most
general allowed by the gauge and supersymmetry transformations that
act at $y = 0, \pi R$. At $y=0$ these transformations are:
\begin{equation}
  \epsilon^{a_+}(x) T^{a_+}, \qquad\qquad 
  \Xi(x,0) =  \left( \begin{array}{c}
		\xi(x) \\
		0
	      \end{array} \right),
\label{eq:y=0sym}
\end{equation}
while, at $y = \pi R$, for the case $W = 1$, the transformations are
\begin{equation}
  \epsilon^{a_+'}(x) T^{a_+'}, \qquad\qquad
  \Xi(x,\pi R) = R(\pi \alpha) \left( \begin{array}{c}
					\xi'(x) \\
					0
				      \end{array} \right).
\label{eq:y=piRsym}
\end{equation}
For $W \neq 1$, the form of the gauge transformations at $y= \pi R$ 
may be more complicated. For $\alpha \neq 0$ the supersymmetries 
on the two branes are different --- for $\alpha = 1/2$ they are orthogonal.

What restrictions are imposed on the brane actions ${\cal L}_4,  
{\cal L}_4'$ by the global symmetries $SU(2)_R$ and $H$? These symmetries 
may be accidental symmetries of the bulk --- a consequence of 5d local
supersymmetry --- so that the brane actions need not respect them.
However, if orbifolding leaves some part of the global symmetry
unbroken, we may choose to impose this on both the bulk and brane actions.

The orbifold transformations ${\cal Z}(0, \pi R)$ may contain
non-trivial contributions from the global symmetries $H$ and
$SU(2)_R$. However, this does not restrict the form of 
${\cal L}_4, {\cal L}_4'$.  It only says that $H$ and/or $SU(2)_R$ 
must be symmetries of the bulk action ${\cal L}_5$.
Of course the brane action at any point can only involve
fields that are even about that point (or derivatives of odd fields). 
However, a brane interaction at $y=0$ can transform non-trivially 
under $Z'$ or, equivalently, $T$. This transformation simply serves 
to fix the brane action at $y = 2\pi n R$ in terms of that at $y=0$ 
--- it does not constrain the action at $y=0$.

We conclude that the global symmetries $H$ and $SU(2)_R$ do not
necessarily place any restrictions on  ${\cal L}_4, {\cal L}_4'$, 
which may be taken to be the most
general set of interactions invariant under the gauge and
supersymmetry transformations at $y = (0,\pi R)$.

\subsection{Calculability}
\label{subsec:calc}

The short distance divergence structure of the theory must reflect
the unbroken local gauge symmetry and supersymmetry. Since the
action was taken as the most general respecting the local symmetries,
all short distance radiative corrections must take the form of local
operators which are already present in Eq.~(\ref{eq:action}).
Thus any quantity which is forced by the local symmetries to vanish at
tree level will have finite UV radiative corrections. Such quantities
need not vanish; they may be generated by IR physics.

As an example, consider the case of $G = SU(3) \times SU(2) \times
U(1)$ with a single Higgs doublet hypermultiplet in the bulk
\cite{Barbieri:2001vh}. The zero-mode structure is precisely that 
of the one Higgs doublet standard model. There has been considerable 
debate recently about the radiative structure of the zero-mode Higgs 
boson mass \cite{Ghilencea:2001ug}, with some arguing that it is 
finite and some that it is quadratically divergent.
The exact unbroken local supersymmetry, given by
Eq.~(\ref{eq:localsusy}) with $\alpha = 1/2$, is sufficient to
guarantee that the mass of the zero-mode Higgs boson is radiatively UV
finite to all orders in perturbation theory. Those who claim divergent
behavior have apparently not realized that there is an unbroken
5d local supersymmetry in the theory.
The Higgs mass is non-zero because at distances larger than $1/R$
there are non-local IR contributions. In the low energy effective
theory (in this case the standard model) these contributions appear to
be quadratically divergent (the usual top loop contribution to the
Higgs mass). However, at shorter distances the locality of the fifth
dimension becomes operative and removes the divergence.
In calculating explicit loop diagrams, this is seen most transparently 
by going to position space for the extra dimension, as in 
Ref.~\cite{Arkani-Hamed:2001mi}.  When using momentum space, the 
internal propagators are expanded in KK modes, and the sum of these 
KK modes must be done in a way which preserves the local supersymmetry. 
A simple way of doing this is to include the contributions from all modes.

Whether a quantity is finite and calculable simply depends on whether
a local operator can be written which contributes to it. In any theory
with the local supersymmetry Eq.~(\ref{eq:localsusy}), all magnetic 
dipole moment operators vanish at tree level and will therefore be
radiatively finite.  Recently a finite one-loop contribution to 
$b \rightarrow s \gamma$ has been computed \cite{Barbieri:2001mr} in the 
theory of Ref.~\cite{Barbieri:2001vh}.  On the other hand the electroweak 
$\rho$ parameter arises at tree level from the supersymmetric operator 
$\delta(y) \int d^4 \theta \; (H^\dagger e^{gV} H)^2$, and hence is 
subject to quadratically divergent radiative corrections.

In calculating radiative corrections to such quantities as the Higgs 
boson mass and $b \rightarrow s \gamma$ one may wonder whether 
contributions from gravitino exchange are important. Since we have 
local supersymmetry in the bulk, such interactions are certainly 
present. They cannot change the above arguments about finiteness, 
and now we argue that they contribute only very small amounts to the 
finite quantities. The local symmetry structure of the theory becomes 
apparent at distances smaller than $1/R$, and hence all contributions 
from shorter distances are cut off.  However, the gravitino 
interactions are weak at scale $1/R$ and do not make a substantial 
contribution.  At some higher energy scale the gravitational 
interaction becomes strong, but these local interactions cannot 
contribute significantly to the finite quantities.

\section{Simple Models with Bulk Higgs}
\label{sec:examples}

In this section we consider simple supersymmetric models with 
$G= SU(5)$ or $G = SU(3) \times SU(2) \times U(1)$, with the Higgs
doublet(s) in the bulk. The breaking of electroweak symmetry is then
linked to the physics of the bulk. In $SU(5)$ theories, a crucial role
of the bulk is to accomplish doublet-triplet splitting. In the
non-unified case, the Higgs is also a near zero-mode. In all cases we
consider the role the bulk plays in breaking supersymmetry.

In general we are interested in non-trivial $U,V,W$ so that $Z$ and
$Z'$ are not simultaneously diagonalized. However, some of the simple
theories do have $[Z,Z']=0$, while in other theories the lack of
commutativity is small, so that it is convenient to think first about
the commuting case.

After global and gauge symmetry breaking, there are a collection of
$H' \times G'$ irreducible hypermultiplets $(\phi, \psi; \phi^c, \psi^c)$. 
Given the gauge and global parities $P_G, P'_G, P_H$ and $P_H'$ 
of Eqs.~(\ref{eq:Z}, \ref{eq:Z'}), the fermion $\psi$ has four 
possibilities for its $(P,P')$ parities: $\psi(p,p')$ with $p,p' = \pm 1$.
The $(P,P')$ parities of all other components of the hypermultiplet are 
now fixed in terms of $P_R \equiv \exp(2 \pi i \alpha \sigma_2) = (+1,-1)$ 
for $\alpha = (0, 1/2)$:
\begin{equation}
  \left[
  \phi(p, P_R p'), \psi(p,p'); \; \phi^c(-p,- P_R p'), \psi^c(-p,-p')
  \right].
\label{eq:hyperpp'}
\end{equation}
There are four different types of hypermultiplet according to whether
$p = \pm$ and $t \equiv pp' = \pm$. If $P_R = +1$ supersymmetry is
unbroken, and there is a zero-mode chiral multiplet only for parities
of equal signs ($t=+1$).  If $P_R = -1$ supersymmetry is broken, and
for parities of equal signs ($t=+1$) the zero-mode is a fermion, 
while for parities of opposite signs ($t=-1$) the zero-mode is a scalar.

Similarly, after gauge symmetry breaking, a gauge boson may have any
combination of parities, $A^\mu(p,p')$, but the other components of
the 5d vector multiplet are then given: 
\begin{equation}
  \left[
  A^\mu(p, p'), \lambda(p, P_R p'); \; 
  (\sigma + i A^5)(-p, -p'), \lambda'(-p, -P_R p')
\right].
\label{eq:vectorpp'}
\end{equation}
The KK mode expansion for any of these fields is given by its $(P,P')$
quantum numbers according to Eq.~(\ref{eq:commuting.eigenfunctions}). 
If $t = +1$ $(-1)$ the KK modes have mass $m_n = n/R$ ($(n + 1/2)/R$). 
The fermion masses of the tower are Dirac type.

It is remarkable that in 5d the most general possible supersymmetry
breaking is described by just a single parameter $\alpha$. For
arbitrary $\alpha$, but keeping $\beta = \gamma =0$, the
eigenfunctions of the $SU(2)_R$ doublets $(\phi, \phi^{c \dagger})$ and
$(\lambda, \lambda')$ pass from Eq.~(\ref{eq:commuting.eigenfunctions}) 
to Eq.~(\ref{eq:N=2eigenfunctions}) with the eigenvalues shifted by 
$\alpha/R$
\begin{equation}
  m_n \rightarrow \left\{ \begin{array}{ll}
			m_n \pm \alpha/R & \;\;\; \mbox{non zero-mode} \\
			\alpha/R         & \;\;\; \mbox{zero-mode}.
		\end{array} \right.
\label{eq:mn}
\end{equation}
The gauginos become Majorana and are shifted in mass relative to their
gauge boson partners. Similarly the hypermultiplet scalars are shifted
in mass relative to their fermionic partners. In both cases the mass
of the zero-mode is lifted by $\alpha/R$, while the excited members of
the $SU(2)_R$ doublets get split in mass by $\pm \alpha/R$ relative to
the corresponding $SU(2)_R$ singlet states.

\subsection{Models with $G= SU(3) \times SU(2) \times U(1)$}

We choose the orbifold symmetries to preserve the gauge group, so that
$P_G$ and $P'_G$ are trivial. All the vector multiplets therefore have
$p=p'=1$ in Eq.~(\ref{eq:vectorpp'}). 

The simplest possibility is that there is a single Higgs
hypermultiplet in the bulk. From Eq.~(\ref{eq:hyperpp'}) we see that if
$t=pp'=1$ for this hypermultiplet, there is a single zero-mode
Higgsino, so that this case is forbidden by anomalies. For $t=-1$ and
supersymmetry unbroken, $R_P=1$, there is no zero-mode Higgs
boson. Such a situation is hard to reconcile with observation: supersymmetry
is unbroken and the Higgs mass squared has a large positive value
comparable to the masses of the KK excitations of the standard model
gauge particles. The unique theory with 1 Higgs hypermultiplet has
$t=-1$ and $\alpha \neq 0$. The case of $\alpha = 1/2$ was studied in
Ref.~\cite{Barbieri:2001vh}. In this theory the Higgs potential depends 
on only 1 unknown parameter, $1/R$, and since the Higgs mass is 
finite it can be predicted: $m_h = 127 \pm 8$ GeV. A deformation of 
this theory is possible by allowing $\alpha$ to deviate from 1/2, 
so that $[Z,Z'] \neq 0$. We study this deformation in section 
\ref{sec:deform}.  The unique 1 Higgs hypermultiplet theory may 
therefore be described by $Z,T$ in the supersymmetry and Higgs flavor 
spaces as
\begin{eqnarray}
  Z &=& \Sigma_3 \otimes 1, 
\label{eq:BHNdef1} \\ 
  T &=& e^{2 \pi i (1/2 + \theta) \sigma_2} \otimes -1.
\label{eq:BHNdef2}
\end{eqnarray}
In many theories it is useful to consider the $Z,T$ basis, since the
symmetry breaking is transparently summarized by $T$. The simplest
assignment of matter is for quark and lepton superfields to all be in
the bulk with positive $T$ parity so that they all contain a single
zero-mode fermion. Thus the orbifold quantum numbers in the matter
flavor space are $(Z_M,T_M) = (+1,+1)$.
Indeed, the requirement that all charged fermions
have Yukawa coupling to the zero-mode Higgs and that the KK modes
would not yet have been discovered makes this all but unique. The only
other possibility known to us has $u_R$ and $d_R$ superfields
located on the branes at $y=0$ and $y= \pi R$, respectively, and the
rest of the matter in the bulk.

The most general theory with two Higgs hypermultiplets is conveniently
described in the supersymmetry and Higgs flavor spaces by
\begin{eqnarray}
  Z &=& \Sigma_3 \otimes \sigma_3, 
\label{eq:2higgs1} \\ 
  T &=& e^{2 \pi i \alpha \sigma_2} \otimes  e^{2 \pi i \beta \sigma_2},
\label{eq:2higgs2}
\end{eqnarray}
and involves two free parameters: $\alpha, \beta$.\footnote{
For $Z = \Sigma_3 \otimes \sigma_0$ the two lightest Higgs modes have 
the same hypercharge.}  
This theory was written down by Pomarol and Quiros \cite{Pomarol:1998sd}, 
who took the view that $\alpha$ and $\beta$ were of order unity. At the 
compactification scale, $1/R$, supersymmetry is broken, so that the 
theory below $1/R$ is non-supersymmetric and must contain a Higgs 
zero-mode. This happens only for the case $\alpha = \beta$, which was 
the focus of their work \cite{Pomarol:1998sd, Delgado:1999qr}. 
Such a light Higgs requires a relation between the breaking of 
supersymmetry, $\alpha$, and the breaking of Peccei-Quinn symmetry, 
$\beta$. We have recently advocated an alternative view 
\cite{Barbieri:2001yz} where $\alpha$ and $\beta$ are taken to
be extremely small. In this case the effective theory below $1/R$ is 
the MSSM. The parameters $\alpha$ and $\beta$ force a non-trivial 
$y$ dependence for the zero-mode Higgs, $h_{u,d}$, and gauginos, 
$\lambda$, so that on compactification they lead to the mass terms
\begin{eqnarray}
  {\cal L} &=& -{\alpha \over 2 R} (\lambda \lambda + \mbox{h.c.})
    - {\alpha^2 \over R^2} \left(h_u^\dagger h_u + h_d^\dagger h_d \right) 
\nonumber \\
  && -{\beta \over R} (\tilde{h}_u \tilde{h}_d + \mbox{h.c.}) 
    - {\beta^2 \over R^2} \left(h_u^\dagger h_u + h_d^\dagger h_d \right) 
\nonumber \\
  && +{2 \alpha \beta \over R^2} (h_u h_d + \mbox{h.c.}).
\label{eq:softops}
\end{eqnarray}
The first line gives the supersymmetry breaking soft masses determined
by $\alpha$ alone, while the second line gives the Peccei-Quinn
breaking terms induced by $\beta$ alone. The third term is
proportional to both supersymmetry and Peccei-Quinn symmetry
breaking. It is remarkable that these mass terms correspond precisely
to those of the MSSM. The common scalar and gaugino mass is
$\alpha/R$, the $\mu$ parameter is $\beta/R$ and the soft parameter
$B$ is predicted to be $2 \alpha/R$. The signs of the two Peccei-Quinn 
breaking terms are correlated such that the conventional $\mu$ parameter 
is negative. It is remarkable that the most general 2 Higgs
hypermultiplet theory in 5d, Eqs.~(\ref{eq:2higgs1}, \ref{eq:2higgs2}), 
leads to the MSSM soft operators, with a unified origin for both 
supersymmetry breaking and the $\mu$ parameter. The smallness of 
supersymmetry breaking and the $\mu$ parameter are both due to the 
smallness of the commutator $[Z,T]$. Quarks and leptons can be on 
either brane, or, if they are in the bulk, they have orbifold quantum 
numbers in the matter flavor space of $(Z_M,T_M) = (+1,+1)$.

\subsection{Models with $G= SU(5)$}

The weakest aspects of conventional 4d grand unified theories 
are the breaking of the unified gauge symmetry and arranging 
the mass splitting of Higgs triplets from Higgs doublets.  
Assigning a non-trivial action for the orbifold symmetries in 
the gauge space, and taking $1/R$ to be the scale of gauge coupling 
unification, opens up new, higher dimensional possibilities 
for grand unification, with orbifold breaking of the gauge group 
and orbifold doublet-triplet splitting.
In the case of 5d, there are two parities $Z$ and $Z'$ available for 
gauge symmetry breaking. If the gauge group is $SO(10)$, there is no 
choice for these parities which gives a set of zero-modes for the 5d 
vector multiplet corresponding to a successful weak mixing
angle prediction. We therefore confine our attention to the case that
the gauge group is $SU(5)$.

To obtain a theory below $1/R$ with (approximate) 4d supersymmetry and
two Higgs doublets, one should start with two Higgs hypermultiplets in
the \textbf{5} of $SU(5)$. The most general orbifold symmetry which 
breaks $SU(5)$ to $SU(3) \times SU(2) \times U(1)$, and does not give
unwanted zero-modes from the 5d vector multiplet, or from the two
Higgs hypermultiplets, is \cite{Barbieri:2001yz}
\begin{eqnarray}
  Z &=& \Sigma_3 \otimes \sigma_3 \otimes I_5, 
\label{eq:2higgsgut1} \\ 
  T &=& e^{2 \pi i \alpha \sigma_2} \otimes  -e^{2 \pi i \beta \sigma_2}
    \otimes \left( \begin{array}{cc} I_3 & 0  \\ 
    0 & -I_2 \end{array} \right),
\label{eq:2higgsgut2}
\end{eqnarray}
where $I_n$ is the $n \times n$ unit matrix.
This theory has unbroken, local 5d supersymmetry transformations of the
form Eq.~(\ref{eq:localsusy}). It also has unbroken, local 5d gauge
transformations. Those corresponding to the generators of the standard
model gauge group, $\epsilon^{3-2-1}(y)$, have $(Z,Z') = (+,+)$, while
the remaining transformation parameters, $\epsilon^X(y)$, have 
$(Z,Z') = (+,-)$.
These transformation parameters therefore have the appropriate KK mode
expansions of Eq.~(\ref{eq:commuting.eigenfunctions}). 
Notice that the full $SU(5)$ gauge transformations are operative at the 
brane at $y=0$, while only those of $SU(3) \times SU(2) \times U(1)$ 
act at the brane at $y = \pi R$. 

In the case that $\alpha = \beta = 0$ the orbifold does not break 4d 
supersymmetry or the Peccei-Quinn symmetry. This is the case
introduced by Kawamura \cite{Kawamura:2000ev} and extended to include 
matter and the unification of gauge couplings \cite{Hall:2001pg}. 
It is important to stress that the orbifold symmetries in Higgs flavor 
space must take the form
\begin{equation}
  (Z_H, T_H) = (\sigma_3, -\sigma_0).
\label{eq:higgsflavor}
\end{equation}
Other assignments do not lead to zero-mode Higgs doublets. For
example, the $\sigma_3$ ensures that the light doublets are
vector-like with respect to the unbroken gauge group. 

Three generations of grand unified matter $(T, \bar{F})$ can be placed
on the brane at $y=0$. The Yukawa couplings to the bulk Higgs fields
are also located at this point, and should therefore be $SU(5)$
invariant.\footnote{
Those of Ref.~\cite{Altarelli:2001qj} are not invariant under the 
$SU(5)$ gauge transformations discussed above.} 
These Yukawa interactions do not induce $d=5$ proton decay, because 
the form of the masses for the Higgs triplets, generated by the 
orbifolding, possesses a symmetry which sets the amplitude to zero 
\cite{Hall:2001pg}. The Yukawa couplings lead to the successful 
$b/ \tau$ mass relation for the third generation. Similar
relations for the lighter generations can be avoided by mixing with
heavy bulk matter \cite{Hall:2001pg}.

Alternatively, matter may be placed in the bulk 
\cite{Hall:2001pg, Hebecker:2001wq}. A single generation requires 
two sets of $\mathbf{10} + \mathbf{\bar{5}}$ hypermultiplets: 
$T, T', \bar{F}, \bar{F}'$.  On the two dimensional space, $(T, T')$ 
and $(\bar{F}, \bar{F}')$, the orbifold symmetry acts as
\begin{equation}
  (Z_M, T_M) = (\sigma_0, \sigma_3),
\label{eq:matterflavor}
\end{equation}
for each generation.
The $\sigma_0$ ensures that the light matter is chiral under the 
unbroken gauge group, while the $\sigma_3$ ensures that an entire 
generation, $q,u,d,l,e$, is massless. Strictly speaking, the 
unification of quarks and leptons is largely lost: 
$T(u,e), T'(q), \bar{F}(d), \bar{F}'(l)$.  However, the $SU(5)$ 
understanding of the quantum numbers of a generation is preserved. 
This combination of bulk matter is the smallest which leads
to anomaly-free, chiral zero-modes, and it automatically gives the 
quantum numbers of a generation.  There is no proton decay from 
$SU(5)$ gauge exchange, and there are no $SU(5)$ fermion mass 
relations \cite{Hall:2001pg}.  Whether matter is placed
on the brane or in the bulk, the theory beneath $1/R$ is the MSSM
without supersymmetry breaking or Peccei-Quinn symmetry breaking.

The only remaining freedom in the structure of the orbifold symmetry
Eqs.~(\ref{eq:2higgsgut1}, \ref{eq:2higgsgut2}) is a non-zero value 
for $\alpha$ and $\beta$ \cite{Barbieri:2001yz}. Since these parameters 
break 4d supersymmetry and the Peccei-Quinn symmetry, they must be 
extremely small.  As in the case of the standard model gauge group, 
they lead to the zero-mode mass terms of Eq.~(\ref{eq:softops}). 

The color triplet Higgsino mass matrix from orbifolding turns off 
dimension 5 proton decay, and bulk matter turns off proton decay from 
the $SU(5)$ gauge interactions and from scalar Higgs triplet 
exchange.  Hence $1/R$ could be reduced to the TeV scale.\footnote{
Grand unified theories at the TeV scale were considered in 
Ref.~\cite{Dienes:1998vh} with a different mechanism of suppressing 
proton decay from the $SU(5)$ gauge interactions.}
While the precise weak mixing angle prediction is lost, 
KK modes of standard model particles, 
as well as those of $X$ and $Y$ gauge bosons and fermions, could be 
produced at high energy colliders. In such grand unified theories 
at the TeV scale, supersymmetry could be broken by the orbifold via 
a large $\alpha$ parameter, and there could be one or two Higgs quasi 
zero-modes.  Above the compactification scale the running of the 
gauge couplings is dominated by $SU(5)$ symmetric power law running 
\cite{Hall:2001pg, Nomura:2001mf}. Thus, for this scheme to be 
viable, there must be some large new exotic contributions to the
gauge couplings either at or beneath the compactification scale.

\section{$SU(3) \times SU(2) \times U(1)$ Model with One Higgs Doublet}
\label{sec:deform}

In this section, we investigate radiative electroweak symmetry breaking 
in one Higgs doublet theories with $G = SU(3) \times SU(2) \times U(1)$.
We consider the case where all three generations of matter and 
a single Higgs propagate in the 5d bulk.  The most general orbifold 
boundary conditions are given by Eqs.~(\ref{eq:BHNdef1}, \ref{eq:BHNdef2}), 
so that we have a 1 parameter family of theories parameterized by a real 
number $\theta$ ($0 \leq \theta < 1/2$).  For any member of this 
family, the Higgs effective potential depends on only one free 
parameter $1/R$, so that the physical Higgs boson mass $m_h$ and 
the compactification scale $1/R$ are calculable.  The $\theta = 0$ 
case corresponds to the model in Ref.~\cite{Barbieri:2001vh}.

The KK mass spectrum for the $\theta = 0$ case is given by
\begin{eqnarray}
  m, h, A^\mu  &:&  n/R
\nonumber\\
  \tilde{m}, \tilde{m}^c, \tilde{h}, \tilde{h}^c, 
  \lambda, \lambda'  &:&  (n+1/2)/R
\\
  m^c, h^c, \sigma  &:&  (n+1)/R,
\nonumber
\end{eqnarray}
where $n = 0,1,2\cdots$ and $m$ represents $q, u, d, l, e$.
A non-zero value for $\theta$ modifies the above mass spectrum such 
that the scalar and gaugino masses are shifted by $\theta/R$.
In particular, the Higgs boson $h$ obtains a tree-level mass 
$\theta/R$, and the two linear combinations of $\tilde{m}$ and 
$\tilde{m}^c$ have split masses of $(n + 1/2 \pm \theta)/R$.
As a result, the Higgs effective potential depends on $\theta$ and 
the values for $m_h$ and $1/R$ also depend on $\theta$.

Radiative electroweak symmetry breaking occurs only when $\theta$ is 
small.  Since the tree-level Higgs mass squared is given by 
$\theta^2/R^2$ and one-loop negative contribution through the top Yukawa 
coupling is $\sim -(1/\pi^4)(y_t^2/R^2)$, $\theta \lsim 1/\pi^2$ is 
required to break electroweak symmetry.  This small $\theta$ perturbs 
the field dependent masses of the top and stop KK towers as
\begin{eqnarray}
  m_{F_n} &=& \left\{ n + \frac{1}{\pi}\, 
    {\rm arctan}(\pi y_t H R) \right\} \frac{1}{R}, \\
  m_{B^\pm_n} &=& \left\{ n \pm \theta + \frac{1}{\pi}\, 
    {\rm arccot}(\pi y_t H R) \right\} \frac{1}{R},
\end{eqnarray}
where $n = -\infty, \cdots, +\infty$, $H \equiv |h|$, 
and there are one Dirac fermion ($F_n$) and two complex 
scalars ($B^\pm_n$) at each level $n$.
With these masses, we can calculate the one-loop Higgs effective 
potential $V_t(H)$ from the top-stop loop, using calculational 
techniques in Ref.~\cite{Antoniadis:1999sd}:
\begin{equation}
  V_t(H) = \frac{9}{16 \pi^6 R^4}
    \sum_{k=1}^{\infty} \left\{ 
    \frac{\cos[ 2k\, {\rm arctan}(\pi y_t R H) ]}{k^5}
    - \cos[2\pi k\theta] \frac{\cos[ 2k\, {\rm arccot}(\pi y_t R H) ]}{k^5}
    \right\}.
\end{equation}
Then, together with the tree-level Higgs potential
\begin{equation}
  V_{H,0}(h) = \frac{\theta^2}{R^2} H^2
    + \frac{g^2 + g^{\prime 2}}{8} H^4,
\end{equation}
we can derive the values for $m_h$ and $1/R$ by requiring that 
$\langle H \rangle = 175~{\rm GeV}$.

\begin{figure}[t]
\centerline{\epsfxsize=9cm \epsfbox{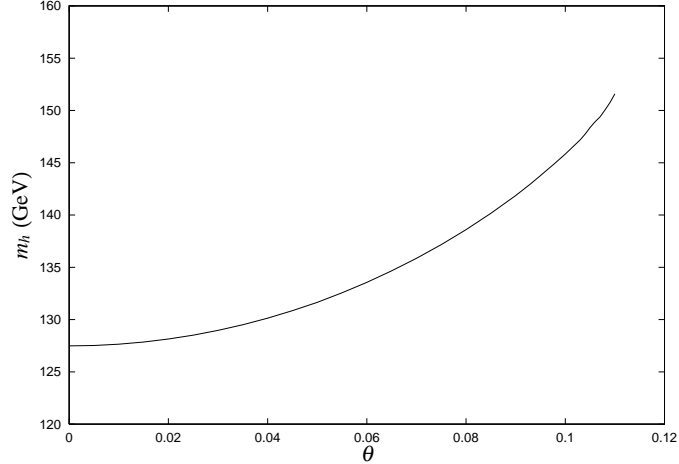}}
\caption{The physical Higgs boson mass $m_h$ as a function of $\theta$.} 
\label{fig:plot-mh}
\end{figure}
\begin{figure}[t]
\centerline{\epsfxsize=9cm \epsfbox{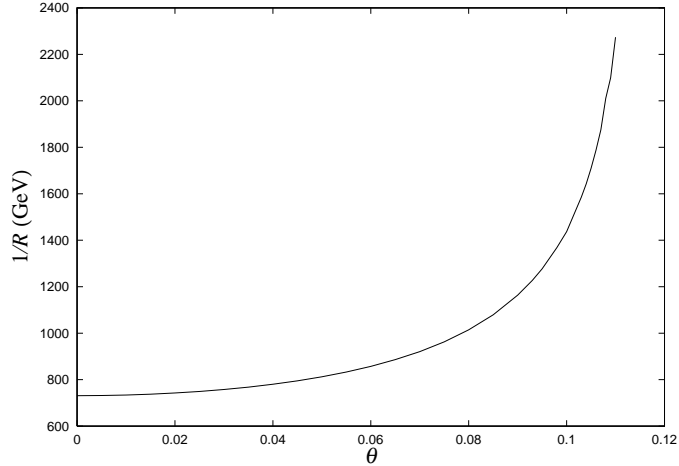}}
\caption{The compactification scale $1/R$ as a function of $\theta$.} 
\label{fig:plot-Mc}
\end{figure}

In Figures \ref{fig:plot-mh} and \ref{fig:plot-Mc}, 
we have plotted the predicted values for 
$m_h$ and $1/R$ as a function of $\theta$, including one-loop gauge 
contributions to the quadratic term in the potential.
They are both monotonically increasing functions with respect to $\theta$.  
The $\theta = 0$ case reproduces the values obtained in 
Ref.~\cite{Barbieri:2001vh}: $m_h = 127~{\rm GeV}$ and 
$1/R = 731~{\rm GeV}$.  Note that the definition of $R$ here is 
different from that in Ref.~\cite{Barbieri:2001vh} by a factor of 2, 
so that it corresponds to $1/R = 366~{\rm GeV}$ in the notation of 
Ref.~\cite{Barbieri:2001vh}.
(A slight increase of $1/R$ compared with the previous value comes 
from an improved treatment of higher order effects.  This also changes 
the central value for the estimate of the lightest stop mass 
to $m_{\tilde{t}_{-}} = 211~{\rm GeV}$.)  As $\theta$ is increased 
to $\theta \gsim 0.1$, $1/R$ approaches infinity meaning that electroweak 
symmetry breaking does not occur beyond that value of $\theta$.
It is interesting that we can obtain larger values for $1/R$ by 
perturbing the model of Ref.~\cite{Barbieri:2001vh}
with small non-zero $\theta$.  It reduces the amount of tuning required 
to obtain phenomenologically acceptable value of the $\rho$ parameter, 
since the contribution from KK towers to the $\rho$ parameter 
scales as $(1/R)^{-2}$.

\section{Conclusions} 
\label{sec:concl}

In this paper we have given the most general form for the orbifold
breaking of symmetries from a single extra dimension. While the
structure of gauge and flavor symmetry breaking is very rich, the
breaking of supersymmetry is described by a single free parameter. The
supersymmetry breaking from all 5d theories is therefore guaranteed to
have a simple form. We have explicitly exhibited the form of the local
supersymmetry transformations which are left unbroken by the
orbifolding. All ultraviolet divergences of the theory must correspond
to local operators which respect this unbroken local supersymmetry. It
is this symmetry that results in the ultraviolet finiteness of the
Higgs mass in certain theories.

We have explicitly constructed the most general orbifold symmetries
for $N=1$ supersymmetric, 5d models with gauge group 
$SU(3) \times SU(2) \times U(1)$ and $SU(5)$, having either one or two
Higgs hypermultiplets in the bulk. There are very few such theories. 
There is a unique one parameter family of $SU(3) \times SU(2) \times U(1)$ 
theories with a single Higgs hypermultiplet. We have studied radiative
electroweak symmetry breaking in this family of theories, and the
Higgs boson mass and the compactification scale are shown as a
function of this parameter, $\theta$, in Figures \ref{fig:plot-mh} and 
\ref{fig:plot-Mc}, respectively.  The special case $\theta = 0$ gives 
a central value for the Higgs mass of 127 GeV \cite{Barbieri:2001vh}.

There is a unique two parameter family of $SU(3) \times SU(2) \times U(1)$ 
theories with two Higgs hypermultiplets. One parameter breaks
supersymmetry and the other breaks Peccei-Quinn symmetry.
This family was first constructed by Pomarol and Quiros 
\cite{Pomarol:1998sd} where the compactification scale was taken to
be in the TeV region. To obtain a zero-mode Higgs boson, 
the two orbifold parameters were taken equal, giving a one dimensional
parameter space. After radiative electroweak
symmetry breaking, the Higgs boson was found to be lighter than 110
GeV throughout this one dimensional parameter space, almost excluding
the theory. Theories with a heavier Higgs boson might result when
the two parameters are allowed to differ by a small perturbation.
Another possibility is that the compactification scale is taken
much larger than the TeV scale, and both parameters are taken very
small \cite{Barbieri:2001yz}. In this case the theory below the
compactification scale is the MSSM with a constrained form for the
soft supersymmetry breaking operators as shown in Eq.~(26). Radiative
electroweak symmetry breaking requires that the compactification scale
be in the interval $10^6 - 10^9$ GeV. 

There is a unique two parameter family of $SU(5)$ theories with 
two Higgs hypermultiplets, where the orbifolding breaks
$SU(5) \rightarrow SU(3) \times SU(2) \times U(1)$. In the special case 
that the two free parameters vanish, the orbifolding corresponds to 
that introduced by Kawamura \cite{Kawamura:2000ev} and developed in 
Ref.~\cite{Hall:2001pg}. The theory including the two free orbifold 
parameters gives a unified origin for $SU(5)$, supersymmetry and 
Peccei-Quinn breaking \cite{Barbieri:2001yz}.

These three families of theories are the only ones in 5d with the stated
gauge groups and bulk Higgs modes. While each family has variants
depending on the location of the quarks and leptons, we have stressed the
tightly constrained form of orbifold symmetry breaking.
In {\it all} cases, the group theoretic structure of the symmetry
breaking is given by Eqs.~(27, 28), where $Z$ is the orbifold parity and 
$T$ the translation symmetry. If there is no $SU(5)$ unification the
last space is removed in these equations with appropriate sign changes 
in the second space. If there is a single Higgs doublet, then in the 
second space $Z$ is +1 and $T$ is $-1$.

\section*{Acknowledgements}

We thank Massimo Porrati, Riccardo Rattazzi and Neal Weiner for 
useful conversations.  Y.N. thanks the Miller Institute for 
Basic Research in Science for financial support.
This work was supported by the ESF under the RTN contract 
HPRN-CT-2000-00148, the Department of Energy under contract 
DE-AC03-76SF00098 and the National Science Foundation under 
contract PHY-95-14797.

\end{document}